%% file: main.tex
\documentclass[10pt,conference]{IEEEtran}

\usepackage{tikz}
\usepackage{amsmath}

\usepackage{booktabs}
\usepackage{filecontents}
\usepackage{marvosym}
\usepackage{threeparttable}
\input{macro}

\IEEEoverridecommandlockouts

\begin{document}

\title{Your Fix Is My Exploit: Enabling Comprehensive DL Library API Fuzzing with
Large Language Models}

\author{
    \IEEEauthorblockN{Kunpeng Zhang\IEEEauthorrefmark{1},
                      Shuai Wang\IEEEauthorrefmark{1}\textsuperscript{\Letter},
                      Jitao Han\IEEEauthorrefmark{2},
                      Xiaogang Zhu\IEEEauthorrefmark{3}, 
                      Xian Li\IEEEauthorrefmark{4},
                      Shaohua Wang\IEEEauthorrefmark{2}\textsuperscript{\Letter}, and 
                      Sheng Wen\IEEEauthorrefmark{4}}
    \IEEEauthorblockA{\IEEEauthorrefmark{1}The Hong Kong University of Science and Technology\\
                      \{zkp0625@outlook.com, shuaiw@cse.ust.hk\}}
    \IEEEauthorblockA{\IEEEauthorrefmark{2}Central University of Finance and Economics\\
                      \{hanjitao1@gmail.com, davidshwang@ieee.org\}}
    \IEEEauthorblockA{\IEEEauthorrefmark{3}The University of Adelaide\\
                      xiaogang.zhu@adelaide.edu.au}
    \IEEEauthorblockA{\IEEEauthorrefmark{4}Swinburne University of Technology\\
                      \{xli1, swen\}@swin.edu.au}
    \thanks{\textsuperscript{\Letter}Corresponding author.}
}

\maketitle

\thispagestyle{plain}
\pagestyle{plain}

\begin{abstract}

    Deep learning (DL) libraries are widely used to form the basis of various AI
    applications in computer vision, natural language processing, and software
    engineering domains. Despite their popularity, DL libraries are known to
    have vulnerabilities, such as buffer overflows, use-after-free, and integer
    overflows, that can be exploited to compromise the security or effectiveness
    of the underlying libraries.
   While traditional fuzzing techniques have been used to find bugs in software,
   they are not well-suited for DL libraries. In general, the complexity of DL
   libraries and the diversity of their APIs make it challenging to test them
   thoroughly. To date, mainstream DL libraries like TensorFlow and PyTorch have
   featured over 1,000 APIs, and the number of APIs is still growing.
   Fuzzing all these APIs is a daunting task, especially when considering the
   complexity of the input data and the diversity of the API usage patterns. 

   Recent advances in large language models (LLMs) have illustrated the high
   potential of LLMs in understanding and synthesizing human-like code.
   Despite their high potential, we find that emerging LLM-based fuzzers
   are less optimal for DL library API fuzzing, given their lack of in-depth
   knowledge on API input edge cases and inefficiency in generating test
   inputs.
   In this paper, we propose \tool, a LLM-driven DL library fuzzing approach. We
   have two key insights: (1) With high \textit{reasoning} ability, LLMs can
   replace human experts to reason edge cases (likely error-triggering inputs)
   from checks in an API's code, and transfer the extracted knowledge to test
   other (new or rarely-tested) APIs. (2) With high \textit{generation} ability,
   LLMs can synthesize initial test programs with high accuracy that automates
   API testing. \tool\ provides LLMs with a novel ``white-box view'' of DL
   library APIs, and therefore, can leverage LLMs' reasoning and generation
   abilities to achieve comprehensive fuzzing. Our experimental results on
   popular DL libraries demonstrate that \tool\ is able to cover more APIs than
   SOTA (LLM-based) fuzzers on TensorFlow and PyTorch, respectively. Moreover,
   \tool\ successfully detected 37 bugs, with 8 already fixed and 19 replicated 
   by the developer but still under investigation.
\end{abstract}

\IEEEpeerreviewmaketitle

\input{introduction}

\input{motivation}
\input{design}
\input{evaluation}

\input{discussion}

\section{Conclusion}

In this paper, we present \tool, a novel white-box approach for LLM-based DL
library API fuzzing. Using LLMs, we infer edge cases and generate initial test
programs, which offer effective and efficient DL library API fuzzing.
Evaluations show that \tool consistently outperforms existing DL library fuzzers
for PyTorch and TensorFlow.


\section*{Acknowledgments}
The HKUST authors are supported in part by a RGC GRF grant under the contract 16214723 and a RGC CRF grant under the contract C6015-23G.

\bibliographystyle{plain}
\bibliography{bib/ref,bib/similarity,bib/decompiler,bib/machine-learning,bib/attack}

\end{document}

%% file: macro.tex
\usepackage{graphicx}
\usepackage[many]{tcolorbox}
\usepackage{mathtools,latexsym,amsfonts,stmaryrd}
\usepackage{pbox}
\usepackage{amsthm}
\usepackage{multirow}
\usepackage{tikz}
\usepackage{mathrsfs}
\usepackage{mathpartir}
\usepackage[ruled,linesnumbered,vlined,noend]{algorithm2e}
\usepackage{url}
\usepackage{syntax}
\usepackage{framed}
\usepackage[noend]{algpseudocode}
\usepackage{flushend}
\usepackage{listings}
\usepackage{moresize}
\usepackage{wrapfig}
 
\usepackage{seqsplit}
\usepackage{alltt}
\usepackage{xspace}
\usepackage[normalem]{ulem}
\usepackage{enumitem}
\usepackage{pifont}
\DeclareMathAlphabet{\mathcal}{OMS}{cmsy}{m}{n}

\input{defs.tex}

\usepackage{thmtools}

\declaretheoremstyle[spaceabove=\topsep,notefont=\normalfont\itshape]{mystyle}

\definecolor{ForestGreen}{RGB}{34,139,34}

\newcommand{\revise}[2]{{\color{red}{\ifx&#1&\else- #1\fi}} {\color{ForestGreen}{\ifx&#2&\else+ #2\fi}}}%
\renewcommand{\revise}[2]{#2}%

\usetikzlibrary{arrows,patterns, decorations.pathreplacing}

\usepackage[colorlinks]{hyperref}   
\AtBeginDocument{%
  \hypersetup{pdfborder={0 0 1}, urlcolor=.}  
}

\definecolor[named]{ACMBlue}{cmyk}{1,0.1,0,0.1}
\definecolor[named]{ACMYellow}{cmyk}{0,0.16,1,0}
\definecolor[named]{ACMOrange}{cmyk}{0,0.42,1,0.01}
\definecolor[named]{ACMRed}{cmyk}{0,0.90,0.86,0}
\definecolor[named]{ACMLightBlue}{cmyk}{0.49,0.01,0,0}
\definecolor[named]{ACMGreen}{cmyk}{0.20,0,1,0.19}
\definecolor[named]{ACMPurple}{cmyk}{0.55,1,0,0.15}
\definecolor[named]{ACMDarkBlue}{cmyk}{1,0.58,0,0.21}
\hypersetup{
  colorlinks,
  linkcolor=ACMPurple,
  citecolor=ACMPurple,
  urlcolor=ACMDarkBlue,
  filecolor=ACMDarkBlue
}

\usepackage{inconsolata} 
\usepackage{biolinum} 

\usepackage{microtype}

\usepackage[noadjust]{cite}

\newcommand{\F}{Fig.}

\newcommand{\T}{Table}
\renewcommand{\S}{Sec.}
\newcommand{\A}{Alg.}

\newcommand{\ignore}[1]{}
\newcommand{\mysubref}[2]{\hyperref[#1]{\ref*{#1}(#2)}}

\lstdefinestyle{base}{
  moredelim=**[is][\color{red}]{@}{@},
  escapeinside={<@}{@>}
}

\usepackage{pifont}

\newcommand\DejaVuttfamily{%
  \fontfamily{DejaVuSansMono-TLF}\selectfont
}

\lstdefinestyle{base}{
  moredelim=**[is][\color{red}]{@}{@},
  escapeinside={<@}{@>}
}

\lstdefinelanguage
   [x64]{Assembler}     
   [x86masm]{Assembler} 
   {morekeywords={CDQE,CQO,CMPSQ,CMPXCHG16B,JRCXZ,LODSQ,MOVSXD, %
                  POPFQ,PUSHFQ,SCASQ,STOSQ,IRETQ,RDTSCP,SWAPGS, %
                  rax,rdx,rcx,rbx,rsi,rdi,rsp,rbp, %
                  r8,r8d,r8w,r8b,r9,r9d,r9w,r9b,reg128,m128}} 

\let\OLDthebibliography\thebibliography
\renewcommand\thebibliography[1]{
  \OLDthebibliography{#1}
  \setlength{\parskip}{0pt}
  \setlength{\itemsep}{1pt plus 0.85ex}
}
\usepackage{listings}
\usepackage{fancyvrb}
\usepackage{color}
\definecolor{lightgray}{rgb}{.9,.9,.9}
\definecolor{darkgray}{rgb}{.4,.4,.4}
\definecolor{purple}{rgb}{0.65, 0.12, 0.82}
\definecolor{commentgreen}{RGB}{63,127,95}
\definecolor{pyblue}{RGB}{59,117,175}
\definecolor{pyorange}{RGB}{239,134,54}
\definecolor{pygreen}{RGB}{81,158,62}

\usepackage{xcolor}
\colorlet{myPurple}{blue!40!red}
\definecolor{myOrange}{RGB}{255,192,0}

\lstdefinelanguage{Solidity}{
  keywords={len,delete,int,void,payable, public, event, contract, typeof, new, true, false, catch, function, return, null, catch, switch, var, if, while, do, else, case, break,struct,const,socklen_t,sa_familty_t,char,sockaddr,load},
  keywordstyle=\color{violet}\bfseries,
  ndkeywords={class, export, boolean, throw, implements, import, this},
  ndkeywordstyle=\color{darkgray}\bfseries,
  identifierstyle=\color{black},
  sensitive=false,
  comment=[l]{//},
  escapeinside={(*@}{@*)},          
  morecomment=[s]{/*}{*/},
  commentstyle=\color{commentgreen}\ttfamily,
  stringstyle=\color{red}\ttfamily,
  morestring=[b]',
  morestring=[b]"
}

\newcommand{\rnum}[1]{\uppercase\expandafter{\romannumeral #1\relax}}

\usepackage{xcolor}

\algnewcommand{\LeftComment}[1]{\Statex \(\triangleright\) #1}

\definecolor{pptbrown}{RGB}{132,60,12}
\definecolor{pptgreen}{RGB}{56,87,35}
\definecolor{pptred}{RGB}{155,30,20}
\definecolor{pptdy}{RGB}{127,96,0}

\newcommand{\parh}[1]{\noindent\textbf{#1}}

\newcommand{\tool}{\textsc{DFuzz}\xspace}

\newcommand{\rom}[1]{\uppercase\expandafter{\romannumeral #1\relax}}

%% file: defs.tex
\usepackage{amsmath, listings, amsthm, proof, xspace}



%% file: introduction.tex
\section{Introduction}
\label{sec:introduction}

To date, deep neural networks (DNNs) and their enabled applications have been
extensively utilized in various real-world scenarios, such as autonomous driving~\cite{liu2020computing, grigorescu2020survey, muhammad2020deep}, software vulnerability detection~\cite{wen2023less,li2023commit,wang2023deepvd,li2021vulnerability}
and medical diagnosis~\cite{shen2017deep, bakator2018deep}. Typically, those DNN applications are built on top of
deep learning (DL) libraries, such as TensorFlow~\cite{tf} and PyTorch~\cite{pytorch}, which offer a
comprehensive set of API functions to facilitate developing DNN models and
execute them. Given that DNN applications have been actively employed in
reliability-sensitive scenarios~\cite{iftikhar2022deep, shafiee2020deep, afshari2023deep}, it is demanding to thoroughly test DL libraries
and uncover underlying bugs in those API functions. Among all popular methods,
fuzzing is deemed the mainstream approach with high potential, given its high
flexibility and capability in detecting real-world defects~\cite{zhu2022fuzzing, bohme2017directed, peng2018t, godefroid2008, godefroid2017learn, pham2019smart}.

While traditional fuzzing techniques have been used to find bugs in software~\cite{zalewski2017american, fioraldi2020afl, lyu2022ems, aschermann2019redqueen, fioraldi2022libafl},
recent works show that they are not well-suited for DL libraries~\cite{titanfuzz, fuzzgpt, IvySyn, wei2022free, gu2022muffin, deng2022fuzzing, xie2022docter, deng2023differential}. In general,
the complexity of DL libraries and the diversity of their APIs make it
challenging to test them thoroughly. To date, mainstream DL libraries like
TensorFlow and PyTorch have featured over 1,000 APIs, and this number
continues to increase. Performing fuzzing on all of these library APIs is a
challenging endeavor, particularly when taking into account the input data
complexity, emerging data types, and diverse API implementation patterns.
Recent DL library fuzzing works primarily rely on manually crafted
mutators~\cite{IvySyn}, which require expensive human efforts and are less
scalable. 

Large language models (LLMs) are transformer-based neural networks that have
achieved state-of-the-art (SoTA) performance in a wide range of natural language
and code processing tasks~\cite{devlin2018bert, wermelinger2023using,
feng2020codebert, brownlanguage, yadavally2024learning, yadavally2023partial,li2022utango, wang2024natural, li2023contextuality, li2021context}. It is shown that LLMs can reason and generate
human-like code, given that they have been trained on large-scale corpora which
often subsume common sense knowledge and programming expertise. Recent
works~\cite{titanfuzz,fuzzgpt} use LLMs to perform DL library API fuzzing with
promising results. However, we find that those emerging LLM-based fuzzers often
suffer from a lack of in-depth knowledge on API input edge cases and
inefficiency in generating test inputs (most test inputs are not helpful). This
is mainly due to the fact that LLMs were merely employed to mutate API inputs
using either scheduling algorithm-based schemes~\cite{titanfuzz} or based on
historical data~\cite{fuzzgpt}. As a result, they essentially treat DL library
API fuzzing as a \textit{black-box problem}, where the underlying API
implementations are \textit{neither} well understood \textit{nor} utilized.


\parh{Our Insights.}~This paper presents a novel and comprehensive DL library
fuzzing approach, by bridging de facto LLMs with a ``white-box view'' of DL
library APIs. We have two key insights: (1) LLMs manifest high \textit{reasoning
ability} and possess pre-knowledge over a considerable number of APIs. They can
reason those API input checks (e.g., \texttt{TORCH\_CHECK} in PyTorch) widely
seen in API low-level code, and infer edge cases (likely error-triggering API inputs)
accordingly. Moreover, these edge cases are transferable to effectively stress other
similar APIs and likely uncover more bugs rapidly. (2) LLMs also manifest high
\textit{generation ability}, in that with proper guidance and feedback, they
can synthesize programs invoking a target API with high accuracy, thus
automating the API testing process with high efficiency. 

With the above insights, we propose \tool, a novel LLM-based fuzzing framework
for DL libraries. \tool\ features a three-step approach to perform effective API
fuzzing, with the usage of LLMs in different steps. \tool\ first employs LLMs to
summarize input checks (e.g., \texttt{TORCH\_CHECK}) found in the low-level code
of a DL library API. \tool\ forms edge cases that can provoke those checks, and
further lift the edge cases to an abstract, context-free form that can be easily
transferred to test other APIs with similar input types. Next, \tool\ uses LLMs
to synthesize initial test programs to invoke a target API, where we form a
feedback-driven process to guide the generation. Third, with edge cases
and initial test programs on hand, \tool\ employs LLMs to synthesize diverse
inputs to fuzz a given API. We further design a set of optimizations to make
\tool\ highly efficient and practical.

We evaluate \tool\ to fuzz two mainstream DL libraries, TensorFlow and PyTorch.
We compare \tool\ with SoTA LLM-based fuzzers, TitanFuzz~\cite{titanfuzz} and
FuzzGPT~\cite{fuzzgpt}, and SOTA type-aware mutation-based fuzzer,
IvySyn~\cite{IvySyn}.
We show that \tool\ achieves higher API coverage with much lower LLM usage. 
In fuzzing PyTorch and TensorFlow, \tool\ uses only 7.12\% and 16.81\% of
TitanFuzz's LLM usage (in initial program generation) respectively, 
yet covers 126 and 125 more APIs than TitanFuzz, respectively. 
Moreover, we show that \tool\ finds 37 bugs in the
latest versions of PyTorch and TensorFlow, with 8 already fixed and 19 replicated by the developer but still under investigation. More than 20 bugs exist in previous versions of
TensorFlow and PyTorch which have been extensively tested by TitanFuzz, FuzzGPT,
and IvySyn, yet none of these fuzzers discovered them.
%
In sum, our contributions are as follows: 

\begin{itemize}
    \item We for the first time advocate a \textit{white-box view} in the
    context of LLM-based DL library API fuzzing. The high reasoning and
    generation abilities of LLMs facilitate inferring edge cases and generating
    test programs, which are essential for comprehensive DL
    library API fuzzing.
    
    \item We present a novel LLM-based fuzzing framework, \tool~\cite{dfuzz}, that implements
        the above insight. \tool\ features a three-step approach to delivering
        fuzzing. \tool\ features a set of design principles and optimizations to
        make it highly efficient and practical.
    \item 
        Our results show that \tool\ can consistently outperform SOTA
        (LLM-based) fuzzers, in terms of API coverage and bug finding. 
        37 bugs have been found in the latest versions of
        TensorFlow and PyTorch, with 24 bugs already existing in
        TensorFlow/PyTorch fuzzed by previous tools.
\end{itemize}

%% file: motivation.tex
\section{Motivation}
\label{sec:motivation}



\subsection{Related Work and Limitations}
\label{subsec:limitation}

Fuzzing DL libraries is a challenging and demanding task faced by the community.
Based on seed differences, existing research on fuzzing DL libraries can be
categorized into two types: model-level and API-level fuzzers~\cite{titanfuzz}.
Model-level fuzzers primarily test APIs related to model construction and
training, treating complete DL models as inputs while simultaneously testing
multiple APIs~\cite{pham2019cradle, gu2022muffin, guo2020audee, liu2023nnsmith,
wang2020deep}. The testing scope of API-level fuzzers encompasses all APIs
within the target DL framework, generating programs for each API and
subsequently applying mutations~\cite{wei2022free, titanfuzz, fuzzgpt, IvySyn,
deng2022fuzzing, xie2022docter, yang2023fuzzing}. 

\tool\ belongs to the latter category, focusing on API-level fuzzing
given its high comprehensiveness over APIs. However, with the high complexity
and diversity of DL library APIs, we find that SoTA methods in this category
feature a rather \textit{opaque} understanding of the underlying API usage
patterns and codebases, even if LLMs may have been employed. To elaborate,
existing DL framework fuzzing methods can be categorized into the following
three types: 

\noindent \ding{172} Scheduling algorithm-based approach, such as
TitanFuzz~\cite{titanfuzz}. Such a method pre-defines basic mutation operators,
and then applies these operators to mutate programs based on a scheduling
algorithm. However, the optimization goal of such methods is often not bug
discovery, but rather metrics like the complexity of generated programs. The
mutation is undirected and lacks guidance to effectively stress APIs.
Importantly, given that there are often thousands of APIs in real-world DL
frameworks, our tentative experiments show that such methods fail to allocate
sufficient testing time to each API, making them hardly comprehensive to test
each API instance. For example, in TitanFuzz's experiments, only 60 seconds were
performed to each API~\cite{titanfuzz}, likely struggling to achieve good results.


\noindent \ding{173} Approach based on predefined input mutators, such as
IvySyn~\cite{IvySyn} and DocTer~\cite{xie2022docter}. These methods require experts to analyze existing bug
codes, summarize features, and thus construct suitable mutators. In comparison
to \ding{172}, mutations based on predefined mutators are more directed and
targeted. However, the construction of mutators is labor-intensive and
time-consuming, and the mutators are often not comprehensive enough to cover all
possible edge cases. In IvySyn, researchers manually analyzed 240 CVEs and
summarized dozens of mutators. However, this relies on heavy expert experience
and human resources for analysis, thus lacking automation and scalability. In DocTer, the tool relies on the provided documentation to extract input constraints for API functions.

\noindent \ding{174} Approach based on historical bug-triggering code, such as
FuzzGPT~\cite{fuzzgpt}. Holistically, history-based methods can be seen as an
enhanced version of \ding{172}, where the mutation is guided by historical bug
cases. Nevertheless, we find that such approaches require collecting a large
number of error codes from the Internet; such error codes may be likely
incomplete and not representative of the entire API input space. Moreover,
there exists a noticeable gap between bug codes and test program generation. Bug
codes often consist of a limited number of statements that are genuinely
pertinent to the bug, whereas other statements are irrelevant and frequently
lead to erroneous program generation. Also, due to the substantial differences
in the syntactic forms and usages of APIs, error-triggering features gathered
from existing bug codes may be likely unusable when fuzzing other APIs. \tool\
achieves notably better results than FuzzGPT, as shown in our evaluation
(\S~\ref{sec:evaluation}).

\subsection{Insights Derived from a White-Box Perspective}
\label{subsec:insight}

Conceptually, we view that the limitations of existing methods can be addressed
by shedding a ``white-box'' perspective on the DL library codebase. In general,
DL frameworks like TensorFlow and PyTorch typically comprise multiple layers of
abstraction, and at the lowest level, native low-level APIs implement specific
operations (e.g., tensor operations). Importantly, these implementations include
extensive checks on input parameters to ensure they meet the expected conditions
of the API. In case of errors, corresponding error messages are generated. We
see that these checks contain a rich set of information about edge cases ---
extreme or special situations of API inputs that are likely to trigger bugs. To
ease understanding, we present a real example below.

\begin{figure}[!htbp]
    \centering
    \vspace{-10pt}
    \includegraphics[width=1.0\linewidth]{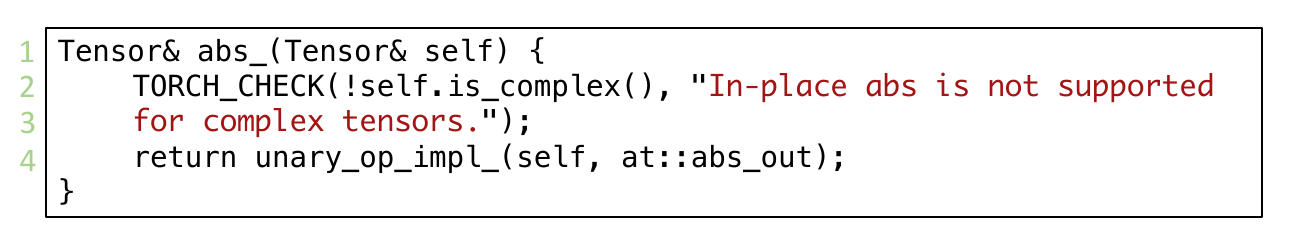}
    \vspace{-20pt}
    \caption{Sample source code in PyTorch.}
    \label{fig:motivation}
\end{figure}

\parh{Real-World Example.}~\F~\ref{fig:motivation} illustrates the low-level
source code of PyTorch API \texttt{torch.Tensor.abs_}. This API computes the
absolute value of a tensor and modifies the tensor in place without creating a
new one. Since \texttt{torch.Tensor.abs_} does not yet support handling tensors
of complex type (i.e., an ``edge case''), line 2 checks whether the input tensor
is complex; if so, an error is thrown.

The check statement (line 2) can be abstracted into an edge case, \textit{a
tensor parameter is a complex tensor}, which likely crashes the API without
the check. Moreover, our manual analysis shows that when testing another API
\texttt{torch.all}, which has the same input parameter type as \texttt{abs_},
feeding the edge case into \texttt{torch.all} triggers an unknown bug. This
example implies that an edge case extracted from one API is often valuable for
testing other APIs. We present the following definition and key observations
that motivate our approach.

\parh{Edge Case.}~Through analyzing the source code of PyTorch and TensorFlow, we
find that the source code of DL frameworks contains a rich set of information
that can be used to infer edge cases. This offers a unique opportunity to leverage
LLMs to extract edge cases from the source code, and then transfer the extracted
knowledge to test a group of APIs with high similarity. \T~\ref{tab:edgeCase}
defines edge cases considered, which are three special conditions over API
inputs. Using such edge cases to test APIs can effectively stress them and
likely uncover defects, if the APIs are not properly handling those edge cases.

\begin{table}[!htbp]
    \centering
    \caption{Edge case categories and examples.}
    \label{tab:edgeCase}
    \resizebox{\linewidth}{!}{
    \begin{tabular}{l | c | c | c}
    \toprule
    Category & Input Example & Expected & Edge Case \\
    \midrule
    Special Type & x is int & x \ is int & x is string \\
    Abnormal Value & x is int & x \textgreater \ 0 & x \textless \ 0 \\
    Special Type Attribute & x is tensor & x.dtype is int & x.dtype is float \\
    \bottomrule
\end{tabular}
    }
\end{table}


%

\parh{Observation I: API Checks Reflect Edge Cases.}~In developing DL
frameworks, developers commonly incorporate check statements to ensure that the
input parameters meet the expected conditions of the API, or alert users if
certain inputs are yet supported. For instance, in the PyTorch low-level code,
we see many check statements, such as \texttt{TORCH_CHECK} and
\texttt{AT\_CHECK}, where each checks one or several edge cases. This suggests
that checks contain rich information to reflect edge cases. 

This paper extracts edge cases by analyzing check statements and then infer the
edge cases based on the context, input types, and error messages. However, in
practice, DL libraries like PyTorch and TensorFlow often have two kinds of check
statements: 1) those encode the edge conditions directly, and 2) those encode
expected conditions of the API inputs. \tool\ indeed considers both types of
check statements to extract edge cases: to extract the former, we directly ask
the LLM to infer the edge case (\S~\ref{subsec:mutation}); to extract the
latter, we ask the LLM to infer an edge case that violates the expected
conditions (by slightly tweaking prompts used in
\S~\ref{subsec:mutation}). To avoid verbosity, we unified the two types of
checks as ``edge cases'' in this paper.

\parh{Observation II: APIs with Same Input Types Often Sharing Edge
Cases.}~Importantly, the edge cases that an API needs to handle are often
determined by their input types. For example, in the case of \texttt{torch.all}
and \texttt{torch.abs_}, where both APIs take a tensor as its input, they share
the same edge cases to handle undefined and empty tensors. 
We view this illustrates an important insight: if APIs have the same input
parameter types, they shall likely share edge cases. Thus, we can transfer
knowledge of edge cases learned from one API to test another API with the same
input types.
Moreover, since edge cases are determined by API input types, we see that
knowledge transfer of edge cases also occurs in a \textit{cross framework
setting}, e.g., an edge case learned from PyTorch can be used to test TensorFlow
APIs, and vice versa. We validate this insight empirically in
\S~\ref{subsec:rq2}; we extract edge cases from PyTorch core libraries and use
these edge cases to test TensorFlow to detect 27 bugs.



\subsection{Pilot Study}
\label{subsec:pilot}

Despite the promising observations, PyTorch/TensorFlow has thousands of APIs,
where each API may contain multiple checks. Extracting the ``edge case'' from a
check statement is not trivial, which requires reasoning the check context,
input types, error messages, and possibly data flow constraints. Relying on
manual analysis would require a significant amount of time and effort, and
certain pattern match (e.g., regular expression) based approaches are also not
feasible.

Having that said, we see the encouraging potential of LLMs in this context. LLMs
are trained on millions of lines of code available on the Internet, and are
shown to manifest promising understanding and reasoning capabilities in relevant
software engineering tasks like code completion, summarization, and bug
fixing~\cite{xu2022systematic, liu2020multi, guo2023longcoder, dinh2024large, li2022utango,pilault2020extractive, zhang2024benchmarking, tang2023evaluating, pearce2023examining, ahmad2024hardware,yin2024thinkrepair,xue2024selfpico,nguyen2019combining}. In fact, we use the code from \F~\ref{fig:motivation} to form a
prompt, and ask GPT-3.5 (ChatGPT)~\cite{gpt35} to extract edge cases from line 2. We find that GPT-3.5
can output the edge case: \textit{``Tensor self is a complex
tensor''} in an accurate and succinct manner. 


This above result is promising, and suggests that LLMs can be used to extract
edge cases from check statements. At this step, we conduct a pilot study to
quantify the feasibility of LLMs in extracting edge cases from DL framework
source code. Without loss of generality, we take PyTorch and GPT-3.5 as
representative examples to validate our approach.
The PyTorch framework comprises front-end, back-end, and bindings. The
front-end, primarily the Python APIs, empowers users to construct and debug DL
models with ease. The back-end, centered around the C++ APIs, implements
elementary operations. Through bindings, PyTorch's front-end interacts with the
C++ implementation in the back-end, blending Python's user-friendly interface
with C++'s high performance. In the back-end, \texttt{Aten}
\texttt{(pytorch/aten/src/ATen)} serves as the underlying tensor library, supporting
tensor operations of various purposes. Most PyTorch high-level APIs and
functionality are built upon \texttt{Aten}, making it a core component of the
entire framework.

We collect Aten's native functions \texttt{(ATen/native)} for our study. Aten's
native functions serve as the contemporary approach for integrating
operators and functions into \texttt{ATen}. 
We randomly select 50 functions from \texttt{ATen/native},
and for each function, we extract its function header and all check statements
(in the form of \texttt{PYTORCH\_CHECK}) to form a code block. 
Then, for each block, we ask GPT-3.5 four questions: (1) \textit{``How many
check statements are contained in each block?''} (2) \textit{``What are the
variables related to each check statement?''} (3) \textit{``What are the types
of variables related to each check statement?''} and (4) \textit{``Extract the
edge cases checked by each check statement.''} We also manually analyze the
selected code blocks to obtain the ground truth. Then, we assess the accuracy of LLM
outputs, whose results are in \T~\ref{tab:llm}.

\begin{table}[!htbp]
    \small
    \footnotesize
    \centering
    \begin{threeparttable}
    \caption{Assessing GPT-3.5 across four tasks to
    comprehend checks.}
    \vspace{-10pt}
    \label{tab:llm}
    \begin{tabular}{l | c }
    \toprule
    Tasks & Accuracy (Success/Total) \\
    \midrule
    The number of checks in a code block & 100\%(50/50\tnote{1} ) \\
    The involved variables of a check & 97.26\%(71/73\tnote{2} ) \\
    The involved variable types of a check  & 97.26\%(71/73) \\
    The edge cases corresponding to a check & 95.89\%(70/73) \\
    \bottomrule
\end{tabular}
\begin{tablenotes}
\item[1] 50: There are a total of 50 code blocks.
\item[2] 73: There are a total of 73 checks among all 50 code blocks.
\end{tablenotes}
\end{threeparttable}
\end{table} 

GPT-3.5 achieves an accuracy of over 95\% on all four tasks. 
LLM can infer the meaning of callee functions based on the context provided. 
Among all 73 \texttt{TORCH_CHECKs}, 54 involve callee functions.
The errors occurred
only in analyzing two code blocks (including three checks). One error occurs
because the target code block contains too much irrelevant information, leading
the LLM to inaccurately analyze it. The other one was misled by the unclear error
message string (\textit{``target(3)''}) in the check. We confirm that without the error message string, the
LLM can correctly analyze the edge case: \textit{`grad_output' is a 3D tensor}.
Overall, we interpret the pilot study as highly promising, and the results
suggest that LLMs can be used to extract edge cases from DL framework source
code with high accuracy.

%% file: design.tex
\section{Design of \tool}
\label{sec:design}

\begin{figure}[!htbp]
    \centering
    \includegraphics[width=1.0\linewidth]{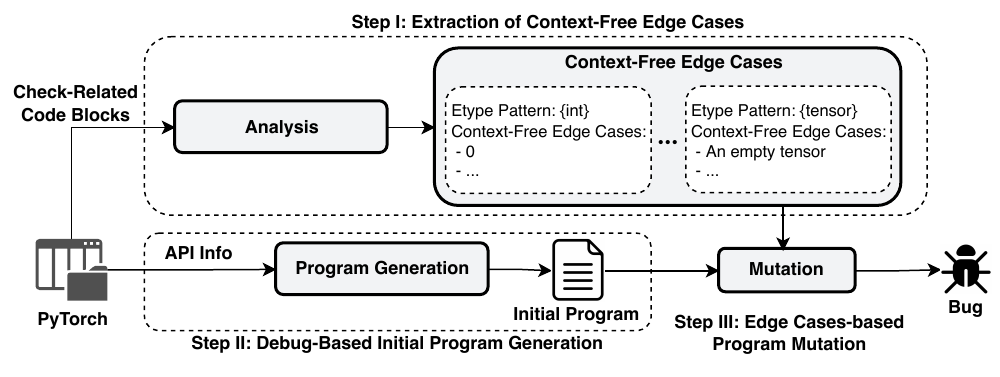}
    \vspace{-20pt}
    \caption{The workflow of \tool.}
    \label{fig:design}
\end{figure}

With the key insights presented in \S~\ref{sec:motivation}, we now present the
design of \tool, a framework for comprehensive API fuzzing. As in
\F~\ref{fig:design}, \tool\ comprises three core steps:

\parh{Step I: Edge Case Extraction.}~Following the discussion in
\S~\ref{sec:motivation}, we use LLM to extract edge cases from the source code
of DL library APIs. We identify and address several domain challenges in this
step (see details in \S~\ref{subsec:extraction}), whose outputs will be edge
cases (conditions over input parameters) in a \textit{context-free} format. This
way, the extracted edge cases can be smoothly used to test other similar APIs,
despite various differences in the input syntactic forms like variable names.

\parh{Step II: Initial Program Generation.}~The next step prepares an initial
test program that invokes a target API for fuzzing. Despite the fact that LLMs
may fail to generate valid programs for certain APIs, we propose an iterative
procedure, where errors encountered during execution are given to the LLM to
regenerate the program and enhance chances of success. 

\parh{Step III: Edge Cases-Based Mutation.}~With edge cases and initial test
programs, \tool\ performs edge cases-centric mutation. Per our observation, we
deem an edge case beneficial for testing a target API, if the input types of the
target API match or subsumes the types of the edge case. With this heuristic, we
instruct the LLM to select beneficial edge cases for fuzzing, achieving high
comprehensiveness in covering even rarely-tested and new APIs as long as their
input types match certain edge cases. Whenever a bug is found, we record the
case and report it to the developers for fixing.

\parh{Application Scope.}~\tool\ is designed for testing DL libraries; it is
fully automated and requires no additional resources besides the source code of
the target DL framework. In contrast, recent works may require input templates
pre-defined by human experts~\cite{IvySyn}, or historical bug reports that are
collected online~\cite{fuzzgpt}. \tool\ can be used by developers and also
normal users who want to test DL libraries (but may lack expertise in library
implementation). While \tool\ is evaluated over two mainstream DL libraries,
PyTorch and TensorFlow, it can be applied to other DL libraries when source code
is available.
\tool\ extracts edge cases from the input checks available in the low-level code
of the DL frameworks. We clarify that such inline checks are the primary way to
handle edge cases in DL libraries. Therefore, analyzing those checks offers a
comprehensive coverage of edge cases. Looking ahead, \tool\ may be extended to
test other types of software, such as those financial software, where inline
checks appear also prevalent~\cite{wang2023verifying}.


\subsection{Edge Case Extraction}
\label{subsec:extraction}

\begin{figure*}[!htbp]
    \centering
    \includegraphics[width=0.75\linewidth]{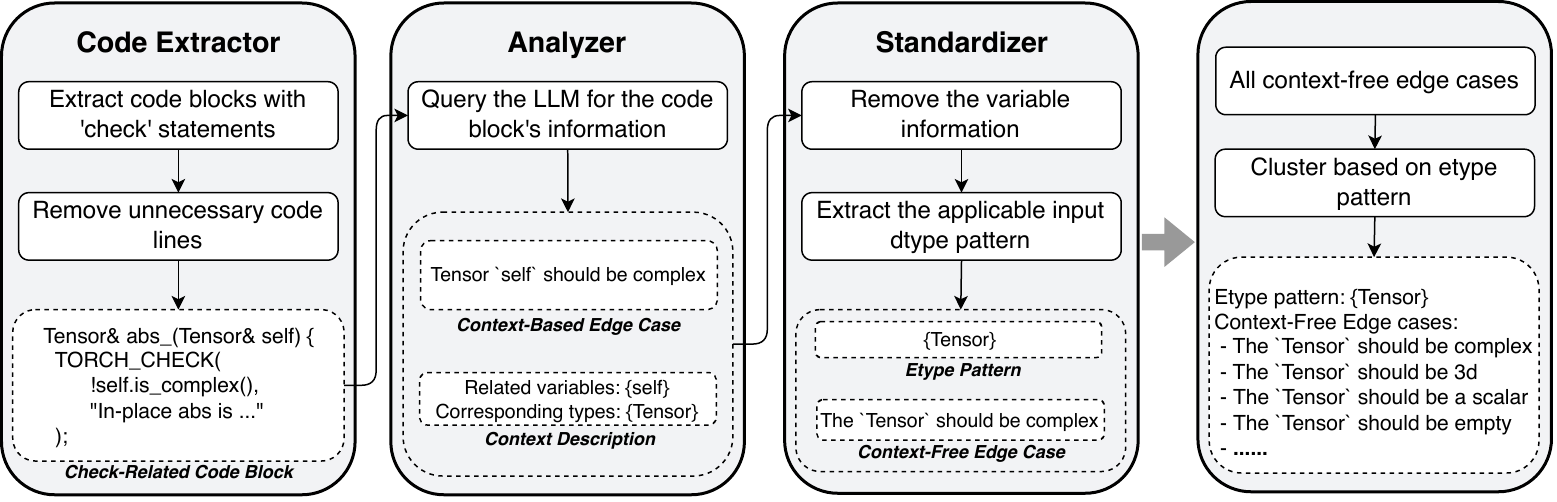}
    \vspace{-5pt}
    \caption{The workflow for extracting edge cases.}
    \label{fig:edgeCase}
\end{figure*}

\tool\ extracts edge cases from the source code of DL libraries and reuses them
to test other APIs. The most straightforward approach is to split the check
statements from the source code and ask LLM to reason edge cases reflected by
these check statements. For instance, we can use prompts like ``\textit{Extract
edge cases detected by the following check statement for fuzzing APIs}'' to
guide LLMs. Nevertheless, our tentative exploration shows the following
challenge.

%
\parh{Transferability.}~LLMs often focus solely on triggering edge cases within
a specific context, without considering whether these extracted edge cases can
be applied to testing other APIs. For instance, in \F~\ref{fig:motivation}, the
edge case extracted from line 2 is \textit{``\texttt{self} is a complex
tensor.''} However, this edge case can only be used when testing a specific API,
\texttt{torch.Tensor.abs_}, due to the reason that other APIs may not have an
input variable named ``\texttt{self}''. Even worse, our analysis focuses on the
low-level implementation of APIs, where variable names may presumably differ
from those at the high level. For example, the variable \texttt{self} at the
high level (Python layer) corresponds to \texttt{Tensor input}. Therefore,
directly utilizing those edge cases extracted by LLMs is not effective.


To address the challenge, we propose a three-step approach to extracting
\textit{context-free} edge cases. We refer the three steps as Code Extractor,
Analyzer, and Standardizer. Code Extractor retrieves \textit{check-related code
blocks} from the DL framework source code, including check statements and the
interface information (function name and input parameters) of their respective
functions. Then, for each check statement within the extracted code block,
Analyzer uses an LLM to extract a \textit{context-based edge case} and the
\textit{context description}; the former refers to the involved input parameters
and their description, and the latter encodes the names and types of the related
parameters. 
%
We further convert context-based edge cases into context-free ones. For each
context-based edge case and its context description, Standardizer first removes
variable names in an edge case, and then uses the data types of the involved
variables in an edge case (referred to as etype pattern) to form a context-free
edge case. After analyzing all check statements, we cluster all (etype pattern,
context-free edge case) tuples; tuples in each cluster share the same etype
pattern.

\subsubsection{Code Extractor}

\begin{figure}[!htbp]
    \centering
    \includegraphics[width=0.9\linewidth]{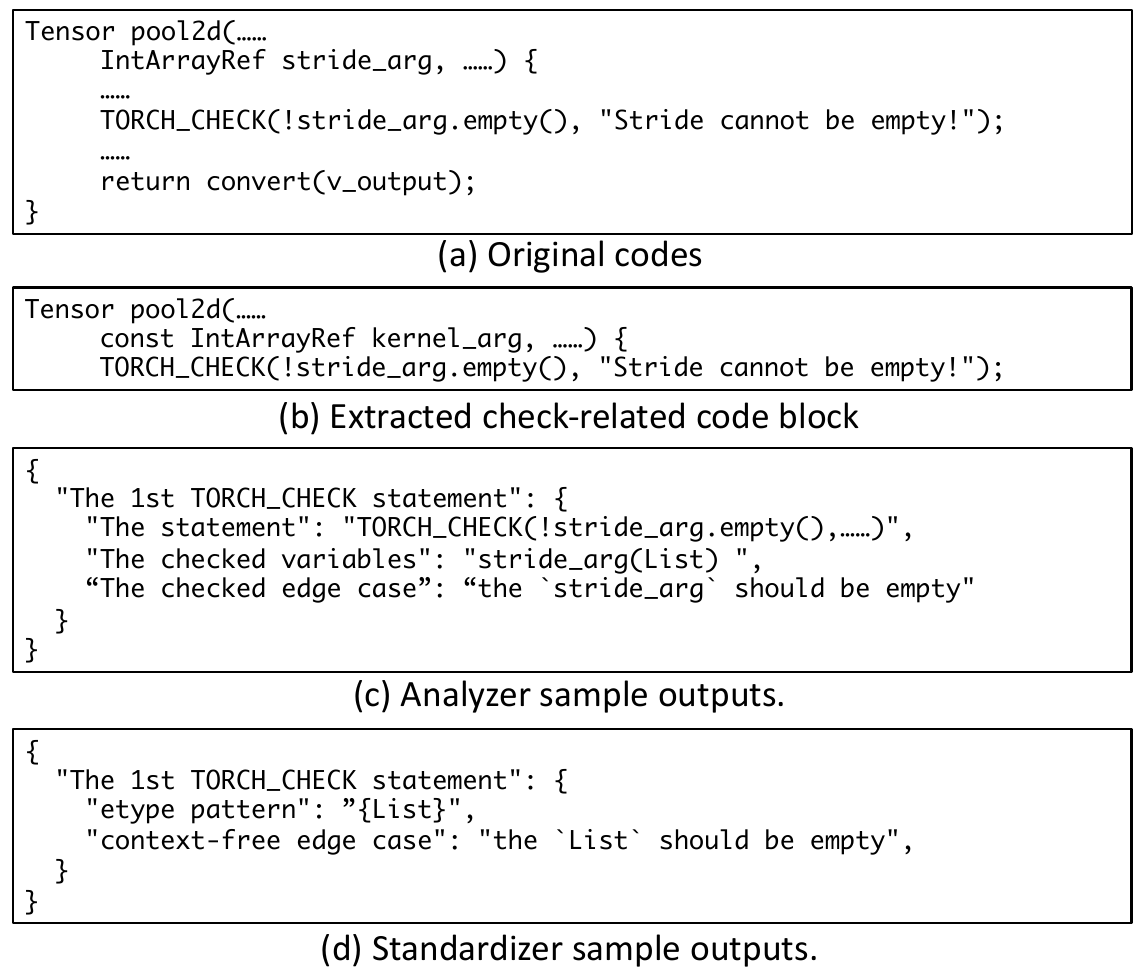}
    \vspace{-10pt}
    \caption{Edge case extraction example.}
    \label{fig:exampleExtractCode}
    \vspace{-5pt}
\end{figure}

Code Extractor extracts code blocks related to edge cases,  whose approach is
generally applicable to various C++ libraries containing check statements. To
ease presentation, consider the ATen library in PyTorch that uses the
\texttt{TORCH\_CHECK} macro to form check statements. We locate all
\texttt{TORCH\_CHECK} statements across source code files under ATen. While
there exists a large number of \texttt{TORCH\_CHECK} statements, we only need to
consider cases where the input parameters of APIs are directly checked by
\texttt{TORCH\_CHECK} statements, given that these checks reflect edge cases
that can be triggered by mutating input parameters.
Furthermore, for each function, we assemble a \textit{check-related code block},
which consists of the function interface and all \texttt{TORCH\_CHECK}
statements within the function. For instance, for the function
\texttt{pool2d} in \F~\ref{fig:exampleExtractCode}(a), the Code
Extractor only retains the function interface and the targeted
\texttt{TORCH_CHECK} statements, resulting in the check-related code block shown
in \F~\ref{fig:exampleExtractCode}(b).

\subsubsection{Analyzer}

After extracting the check-related code blocks, Analyzer extracts the
context-based edge case for each check statement. To do so, one approach is to
directly employ LLMs to analyze each check statement and determine its
corresponding edge case. Though it is feasible, the analysis results are
challenging to use subsequently because: (1) The randomness of the output format
makes it difficult to extract information in a unified manner. For instance, our
employed LLM (GPT-3.5) sometimes provides edge cases directly corresponding to
each check, while at other times, it divides the analysis outputs into multiple
paragraphs, making our subsequent analysis tedious. (2) Lacking relevant
variable information makes it difficult to convert context-based edge cases into
context-free forms. For example, the edge case extracted by LLM for
\F~\ref{fig:exampleExtractCode}(b) is \textit{the ``stride_arg'' should be
empty.} Yet, for other APIs, it is unclear what \textit{``stride_arg''} refers
to. 

\begin{figure*}[!htbp]
    \centering
    \includegraphics[width=0.85\linewidth]{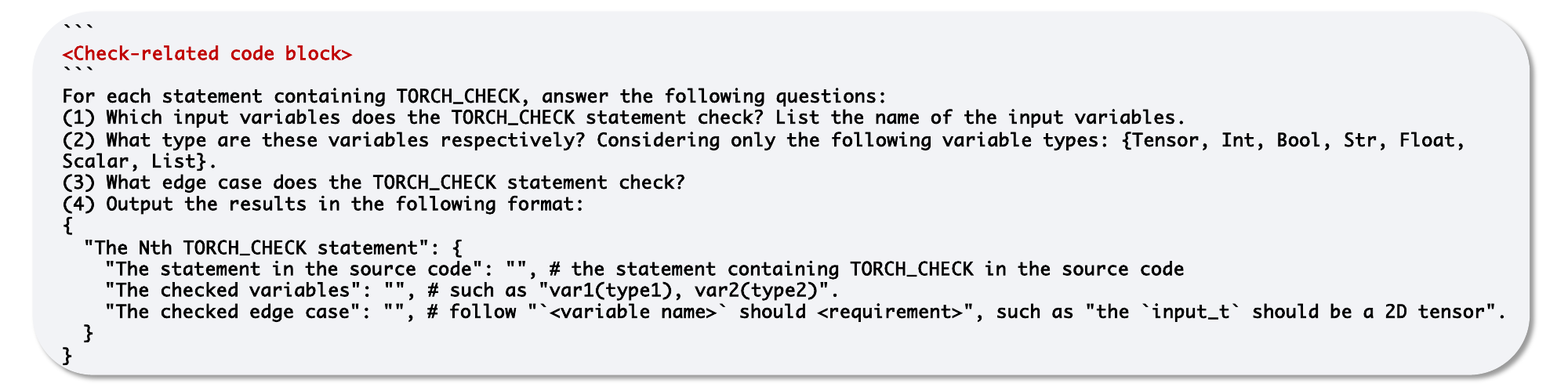}
    \vspace{-10pt}
    \caption{The prompt used by Analyzer.}
    \label{fig:promptAnalyzer}
\end{figure*}

\parh{Prompt Design.}~To form context-free edge cases, we instruct the LLM to
reason edge cases and the context description (variable names and types)
associated with each check. The prompt is shown in \F~\ref{fig:promptAnalyzer}.
We divide the analysis of a \texttt{TORCH\_CHECK} into four steps:
1. \textit{``What variables does the \texttt{TORCH\_CHECK} examine?''}
2. \textit{``What are the data types of these variables?''} Here, we consider
seven most commonly used types: (\texttt{Tensor}, \texttt{Int}, \texttt{Bool},
\texttt{Str}, \texttt{Float}, \texttt{Scalar}, \texttt{List}), but it is easy to
extend to other types.
3. \textit{``What edge cases does the \texttt{TORCH\_CHECK} check?''}
4. \textit{``To standardize the output and reduce irrelevant information,
summarize the output in JSON format.''} After each JSON item, we also provide
the expected output format and examples.
As an example, for the check-related block in
\F~\ref{fig:exampleExtractCode}(b), the analysis results are in
\F~\ref{fig:exampleExtractCode}(c).

\subsubsection{Standardizer}

Analyzer uses LLM to obtain the context-based edge cases. However, as previously
noted, context-based edge cases are hardly useful. Many context-based edge cases
are redundant, and since variable names remain in an edge case, they can only be
used in specific syntactic contexts. Additionally, these context-based edge cases
are extracted from low-level code, whereas our API fuzzing will be launched at
the Python layer (noted in \S~\ref{subsec:insight}). 

Analyzer has prepared sufficient information for Standardizer to convert
context-based edge cases in a format usable at the Python layer, i.e.,
context-free edge case. In particular, for each context-based edge case, we
gather the types of relevant variables into a set, called the \textit{etype
pattern}. Etype pattern reflects the types of input parameters that an API
expects in accordance with the analyzed edge case. For example, suppose an edge
case's etype pattern is \texttt{\{tensor, tensor\}}, then this edge case can be
used to test APIs that expect two tensor inputs, such as
\texttt{torch.add(input: Tensor, other: Tensor)}. And to obtain context-free
edge cases, we replace all variable names in the context-based edge cases with
the corresponding types. For example, for the context-based edge cases and the
context descriptions extracted from \texttt{pool2d} in
\F~\ref{fig:exampleExtractCode}(b), we present the extracted etype pattern and
context-free edge cases in \F~\ref{fig:exampleExtractCode}(d). Finally, for all
(etype pattern, context-free edge cases) tuples, we eliminate duplicates and
cluster them based on etype pattern.


\subsection{Initial Program Generation}
\label{subsec:generation}

To fuzz a target API $i$, we need to generate a test program that invokes $i$.
The test program should be non-trivial, i.e., it can pass input values to the
target API and check the output. With the high generation ability, we employ
LLMs to synthesize the program, thus alleviating the burden of manually writing
test programs. However, our tentative exploration implies that certain APIs are
presumably not covered by LLM training data, especially for newly added or less
commonly used APIs. This raises a hurdle for LLM to generate valid test
programs.


\begin{figure}[!htbp]
    \centering
    \includegraphics[width=0.9\linewidth]{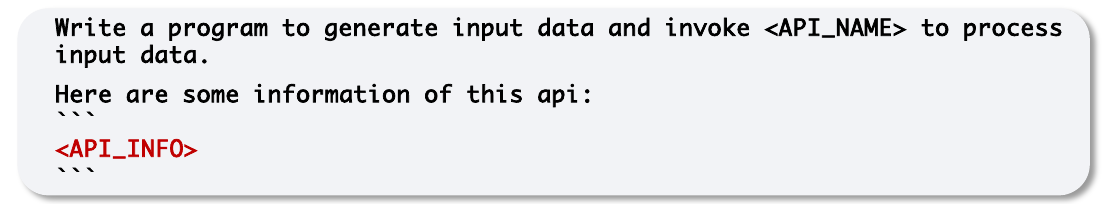}
    \vspace{-10pt}
    \caption{The prompt of initial program generation.}
    \vspace{-5pt}
    \label{fig:promptGeneration}
\end{figure}

To tackle this challenge, we divide the initial program generation into two
stages: the generation stage and the debug stage. In the generation stage, we
integrate the interface information of the target API into a prompt that is fed
into the LLM to synthesize the initial test program. The prompt design is shown
in \F~\ref{fig:promptGeneration}. 
However, per our observation, generating a program that is both syntactically
and semantically correct is challenging; the LLM-generated programs are often
not executable. Therefore, we propose a debug stage, where we instruct the LLM
to reflect the error message and regenerate the program. 

As shown in \A~\ref{alg:programGeneration}, \tool first generates an initial
program based on the interface information of the API and runs it to obtain the
execution status (lines 3--7). If there is an error, we extract the error
function along with the specific error message (line 12). Then, we feed this
error information into the LLM to regenerate a program that can resolve the
current error (lines 12--15). 
If debugging DEBUG_MAX times still fails to obtain a valid program, we ask LLM
to regenerate a new initial program (back to line 5). This is important, as due
to the stochastic nature of LLM, the same prompt can yield programs of different
quality, helping us to avoid getting stuck in ``local minima'' (lines 9--17).
Based on our preliminary exploration, we set INIT_MAX to 2 and DEBUG_MAX to 3 to
balance the trade-off between the quality of the generated program and the time
cost. See our evaluation in \S~\ref{subsec:rq1}.

\begin{algorithm}[!t]
\scriptsize
\caption{Initial Program Generation Algorithm}\label{alg:programGeneration}
\KwIn{The target API, $api$. The interface information of the target API, $api\_info$.}
\KwOut{A valid program that invokes $api$, $P$.}
$init\_cnt \gets 0$\\
\While{$init\_cnt < INIT\_MAX$}{
    $dialogue \gets [\ ]$\\
    $prompt \gets construct\_prompt(api, api\_info)$\\
    $dialogue.append(prompt)$\\
    $P \gets LLM(dialogue)$\\
    $status, msg \gets exec(P)$\\
    $debug\_cnt \gets 0$\\
    \While{$debug\_cnt < DEBUG\_MAX$}{
        \If{$status == SUCCESS$}{
            $ruturn\ P$
        }
    
        $error\_info \gets get\_msg(msg)$\\
        $dialogue.append(P)$\\
        $dialogue.append(error\_info + ``Regenerate")$\\
        $P \gets LLM(dialogue)$\\
        $status, msg \gets exec(P)$\\
        $debug\_cnt \gets debug\_cnt + 1$\\
    }
    $init\_cnt \gets init\_cnt + 1$\\
}
\end{algorithm}

\subsection{Edge Case-Based Mutation}
\label{subsec:mutation}

Through the previous steps, we have obtained the context-free edge cases and the
initial program to invoke a target API $i$. 
To determine which context-free edge cases are presumably beneficial for testing
$i$, we collect the etype pattern of $i$, denoting a combinatorial set of its
parameter types. For example, for \texttt{torch.add(input: Tensor, other:
Tensor)}, its etype pattern is \texttt{\{Tensor, Tensor\}}. Then, we iterate
each subset of the etype: \texttt{\{\}}, \texttt{\{Tensor\}},
\texttt{\{Tensor\}}, \texttt{\{Tensor, Tensor\}}, and search for all collected
context-free edge cases to find a match. Once matched, we concretize the type
names in the context-free edge cases with the names of the input parameters in
the target API. For example, when testing \texttt{torch.all(input: Tensor)}, we
then transform ``\textit{`Tensor' is a complex tensor}'' to ``\textit{`input' is
a complex tensor}''.

\begin{figure}[!htbp]
    \centering
    \includegraphics[width=0.98\linewidth]{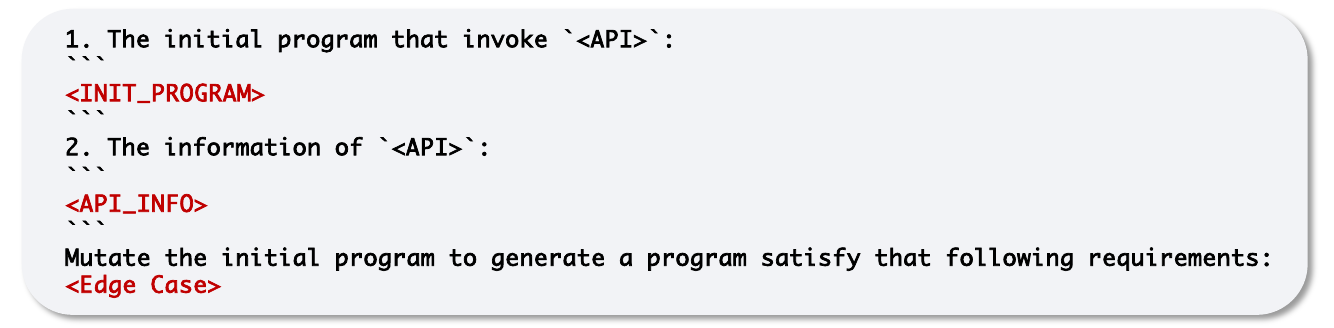}
    \vspace{-10pt}
    \caption{The prompt of mutation.}
    \label{fig:promptMutation}
    \vspace{-10pt}
\end{figure}

Finally, we assemble the collected context-free edge cases, the definition of
the target API $i$, and the corresponding initial program that invokes $i$ into a
prompt; a sample is in \F~\ref{fig:promptMutation}. Then, we let the LLM
generate a test program capable of triggering the edge cases, to see if it can
uncover any bugs. 

As shown in \A~\ref{alg:programMutation}, \tool begins by retrieving the initial
program and etype pattern of the target API (lines 2--3). Then, leveraging the
etype pattern, \tool identifies the context-free edge cases applicable to the
API (line 4). To reduce overhead, we select a subset of matched edge cases for
mutation (line 5). More specifically, we categorize edge cases into two types
based on the etype pattern: individual edge cases and compound edge cases. The
former's etype pattern is only related to a single variable, while the latter
involves multiple variables. In selecting individual cases, we prioritize
variables that appear earlier in the API as they are considered more crucial.
Particularly, for the first two variables, all individual edge cases related to
them will be selected. 25\% of the individual edge cases related to variables in
positions 3--4 of the API will be randomly selected, while the remaining
variables will be chosen at a rate of 12.5\%. Due to the relatively limited
quantity of compound edge cases, all matched compound edge cases will be
selected for mutation. Finally, any bugs triggered by the mutated program are
then collected for further analysis (lines 9--10). Following the convention in this
field~\cite{titanfuzz, fuzzgpt, wei2022free}, we consider two types of bugs:
crashes and CPU/GPU inconsistency. The former asserts that fuzzing triggers any
``crashes,'' including aborts, segmentation faults, etc. For the latter, we run
mutated programs separately on the CPU and GPU to check if the outputs are
consistent.


\begin{algorithm}[!t]
\scriptsize
\caption{Edge Case-Based Mutation Algorithm}\label{alg:programMutation}
\KwIn{The target API, $api$. A library of context-free edge cases, $CEC$.
}
\KwOut{A set of programs that trigger bugs, $BUG$.}
$Bugs \gets [\ ]$\\
$P_i \gets get\_init\_program(api)$\\
$etype\_pattern \gets get\_etypes(api)$\\
$matched\_edge\_cases \gets match(etype\_pattern, CEC)$\\
$selected\_edge\_cases \gets select(matched\_edge\_cases)$\\
\For{$edge\_cas$ in $selected\_edge\_cases$}{
    $P_m \gets mutation(P_i, edge\_case)$ \Comment{Based on Figure~\ref{fig:promptMutation}}\\ 
    $status, msg \gets exec(P_m)$\\
    \If{$status == BUG$}{
        $Bugs.append(P_m)$\\
    }
}
\end{algorithm}

%% file: evaluation.tex
\section{Implementation}
\label{sec:implementation}

The core testing process of \tool\ is implemented in Python with approximately
4,000 lines of code; see our artifact at~\cite{dfuzz}. 

\parh{Check Identification.}~Without losing generality, we extract edge cases
from the checks found in the source code of PyTorch. We see that PyTorch inline
checks are well-structured, in the format of \texttt{TORCH\_CHECK}, making it
easier to analyze and validate. Also, PyTorch updates frequently, resulting in
considerably more checks than other DL libraries on the market. 

Per our investigation, PyTorch's tensor operation library, Aten, contains a
substantial number of checks over different API input types and attributes. This
is reasonable: Aten is one core library of PyTorch that has been developed and
constantly updated by the PyTorch team for decades. Aten contains a wide
range of computation operations, such as matrix multiplication, convolution, and
activation functions. These operations are the foundation of PyTorch's
high-level APIs, such as \texttt{torch.nn} and \texttt{torch.optim}. Thus, we
decide to use PyTorch's ATen library (\texttt{pytorch/aten/src/ATen/native}) to
extract edge cases. We find that using this library offers a good balance
between the number of checks (large enough to cover a wide range of APIs) and
the complexity of the code (not too complex to analyze).

We extracted a total of 20 kinds of etype patterns and their associated 132 edge
cases. The extracted etype patterns include not only 7 basic types
(\texttt{Tensor}, \texttt{Int}, \texttt{Bool}, \texttt{Str}, \texttt{Float},
\texttt{Scalar}, \texttt{List}), but also 13 compound types composed of multiple
basic types, such as (\texttt{Tensor_1}, \texttt{Tensor_2}), (\texttt{Int_1},
\texttt{Int_2}), and so on. In compound types, there are often constraints
between variables, such as \textit{Tensor_1 has a larger last dimension than
Tensor_2} or \textit{Both integers Int_1 and Int_2 are negative}. 
With these extracted context-free edge cases, we conducted tests on the APIs of
PyTorch in \S~\ref{sec:evaluation}. Moreover, it is important to note that the
extracted types and edge cases are transferable to other DL libraries; we also
test them on TensorFlow in \S~\ref{sec:evaluation}.

\section{evaluation}
\label{sec:evaluation}

We aim to answer the following research questions: \textbf{RQ1}: How does \tool\
perform in terms of API coverage? \textbf{RQ2}: How many (new) bugs can be
discovered by the \tool? and \textbf{RQ3}: how effective is \tool\ when using
alternative LLMs? We also conduct an ablation study in \textbf{RQ4}.

\parh{DL Libraries.}~We test two mainstream DL libraries, PyTorch and
TensorFlow, for the following two reasons: (1) PyTorch and TensorFlow are the
most widely-used DL libraries, and bugs discovered within them hold greater
value. (2) PyTorch and TensorFlow have already been tested by many fuzzing
tools. If we can still uncover long-standing bugs from them, it convincingly
shows the effectiveness of \tool. As a fair comparison, we use a consistent set
of target APIs with TitanFuzz and FuzzGPT. In \textbf{RQ1}, we test API coverage
using the same versions of PyTorch (v1.12) and TensorFlow (v2.10) as those in
the TitanFuzz and FuzzGPT papers. In \textbf{RQ2}, we conduct tests on the
latest versions of PyTorch (v2.2.1) and TensorFlow (v2.15) to uncover 
previously unknown bugs.

\parh{Environment.}~All evaluations are conducted on a system with 256 GB RAM
and running Ubuntu 20.04 with one NVIDIA GeForce RTX 3090 GPU.

\parh{LLMs.}~The current prototype of \tool\ uses
GPT-3.5 (\texttt{gpt-3.5-turbo-1106}) for various LLM-based reasoning and generation
tasks. We set the temperature to 0 to get the best results.
In \textbf{RQ3}, we study the feasibility of replacing GPT-3.5 with
llama2-7B-chat and llama2-70B-chat~\cite{llama}; further discussions are in
\S~\ref{sec:discussion}.

\parh{Fuzzers.}~When conducting \textbf{RQ1}, we compare \tool\ with SoTA tools,
TitanFuzz and FuzzGPT. They are both LLM-based fuzzers and have successfully
tested a wide-range of APIs on PyTorch and TensorFlow. For bug discovery
(\textbf{RQ2}), we compare \tool\ with TitanFuzz, FuzzGPT, and IvySyn. IvySyn is
another SoTA fuzzing tool for DL frameworks. We have introduced the design of
these SoTA tools in \S~\ref{subsec:limitation}. In line with these previous
tools, we mainly use two metrics, (1) API coverage (\textbf{RQ1}) and (2) number
of detected bugs (\textbf{RQ2}), to assess \tool\ and compare with previous
tools.

\subsection{RQ1: API Coverage}
\label{subsec:rq1}

API coverage is an important metric for DL library fuzzing, given that a bug is
triggered only when the related APIs are invoked~\cite{titanfuzz,
deng2022fuzzing, wei2022free, yang2023fuzzing}. Evaluating API coverage reflects
the quality of the initial programs generated by \tool. We first report API
coverage below, then measure the debugging procedure of \tool\ when synthesizing
these initial programs.

We compare \tool\ with FuzzGPT and TitanFuzz, both of which are LLM-based
fuzzers. IvySyn focuses on testing low-level, native DL (C/C++) code. It synthesizes code snippets in high-level languages (e.g., Python) only for the buggy code identified in the low-level code. Consequently, IvySyn's approach isn't optimized for API coverage, and therefore, we do not compare it.
As a fair comparison, we conduct experiments on PyTorch (v1.12) and
TensorFlow (v2.10) (the same setup used in the FuzzGPT and TitanFuzz papers).
More specifically, we conduct experiments directly within the Docker environment
provided by TitanFuzz. We use the default settings of TitanFuzz and FuzzGPT. 

\begin{table}[!htbp]
\centering
\caption{The comparison of API coverage. PyTorch has a total of 1,593
APIs, while TensorFlow has 3,316 APIs.}
\vspace{-5pt}
\resizebox{\linewidth}{!}{
\begin{threeparttable}
\begin{tabular}{@{\extracolsep{4pt}}lccccccccccc@{}}
\toprule
\multirow{2}{*}{\textbf{Target Library}} & \multicolumn{2}{c}{\textbf{\tool}} & \multicolumn{2}{c}{\textbf{FuzzGPT}} & \multicolumn{2}{c}{\textbf{TitanFuzz}} & \textbf{Under}\\
\cline{2-3} \cline{4-5} \cline{6-7}

& Cov\tnote{1} & Times\tnote{2} & Cov & Times  & Cov & Times & \textbf{Test}\\
\midrule
PyTorch(v1.12) & \textbf{1455} & \textbf{2839} & 1377 & Unknown & 1329 & 39825 & 1593\\
TensorFlow(v2.10) & \textbf{2340} & \textbf{13938} & 2309 & Unknown & 2215 & 82900 & 3316\\

\bottomrule
\end{tabular}
\begin{tablenotes}[para]
\item[1]Cov: API coverage.
\item[2]Times: Number of times LLM has been invoked.
\end{tablenotes}
\end{threeparttable}
}
\vspace{-5pt}
\label{tab:coverage}
\end{table}

We measure the number of covered APIs and the number of LLM invocation times for
each tool, whose results are in \T~\ref{tab:coverage}. \tool and TitanFuzz both
comprise two stages: initial program generation and mutation. As the mutation
phase does not lead to additional gains in API coverage, we collect API coverage
and LLM usage times from the initial program generation stage for both these
fuzzers. For FuzzGPT, we directly use its final coverage data. \tool\ offers the
highest API coverage in both PyTorch and TensorFlow. In particular, \tool\
covers 78 and 126 more APIs on PyTorch compared to FuzzGPT and TitanFuzz,
respectively. Meanwhile, \tool\ invokes significantly less amount of LLMs than
that of TitanFuzz. FuzzGPT has not been open-sourced at the time of writing, and
its LLM usage counts are unknown to us. On PyTorch and TensorFlow, the LLM usage
of \tool is only 7.12\% and 16.81\% of TitanFuzz's, respectively. We believe
the results are promising, given that \tool\ is able to cover more APIs with
such fewer LLM invocations. 

Recall our synthesis follows a trial-and-fix strategy where a synthesis process
deems failed if \tool\ cannot successfully debug an initial program within
a certain number of attempts (\A~\ref{alg:programGeneration}). Here, we
report the number of successful and failed debugging attempts when generating
the initial programs in \T~\ref{tab:debug}. Note that successful debug refers to
the scenario where the program first synthesized by the LLM is
\textit{invalid}, yet, after debugging, a valid program is obtained; ``failed
debug'' has been noted above.

\begin{table}[!htbp]
    \vspace{-5pt}
    \centering
    \caption{Debug success evaluation.}
    \vspace{-5pt}
    \label{tab:debug}
    \resizebox{0.8\linewidth}{!}{
    \begin{tabular}{c | c | c | c   }
    \toprule
    Target Library & Successful debug & Failed debug & Success rate \\
    \midrule
    PyTorch & 198 & 138 & 58.92\% \\ 
    TensorFlow & 902 & 976 &  48.02\%\\ 
    \bottomrule
\end{tabular}
    }
    \vspace{-5pt}
\end{table} 

Overall, \tool's debugging success rates on PyTorch and TensorFlow are 58.92\%
and 48.02\%, respectively. PyTorch has a total of 1,593 APIs, where for
the 1,257 APIs, \tool successfully generates valid programs with just one query
to the LLM, and the remaining 336 APIs require debugging. Among these, 198 APIs
yield corresponding invocation programs after debugging, while the remaining 138
APIs cannot be debugged successfully. We report that debugging each PyTorch API
requires invoking the LLM approximately $4.71$ times.
With further analysis, we find that the gap between PyTorch and TensorFlow is
due to the complexity of the APIs. In particular, there are more rarely-used
APIs in TensorFlow, for which our employed LLM appears to lack related
knowledge. This makes the overall generation and debugging process more
challenging compared to PyTorch. Overall, we believe the results are promising,
indicating that when integrating error information into prompts, we effectively
guide the LLM to generate valid programs.

\begin{table*}[!htbp]
    \centering
    \caption{The Bugs found by \tool.}
    \vspace{-5pt}
    \label{tab:bugs}
    \resizebox{0.7\linewidth}{!}{
    \begin{tabular}{c | c | c | c | c | c }
    \toprule
    \multirow{2}{*}{Target Library} & \multirow{2}{*}{Total} & \multirow{2}{*}{Fixed} & \multirow{2}{*}{Replicated} & Existed in PyTorch (v1.12) / TF (v2.10) & Existed in PyTorch (v1.11) / TF (v2.6)\\
    &  &  & & (Not found by TitanFuzz/FuzzGPT) & (Not found by IvySyn)\\
    \midrule
    PyTorch & 10 & 4 & 4 & 8 & 6\\
    TensorFlow & 27 & 4 & 15 & 16 & 16\\
    \midrule
    Total & 37 & 8 & 19 & 24 & 22\\
    \bottomrule
\end{tabular}
    }
\end{table*} 

\begin{table}[!htbp]
    \centering
    \caption{Bug types.}
    \label{tab:bugTypes}
    \vspace{-5pt}
    \resizebox{0.85\linewidth}{!}{
    \begin{tabular}{c | c | c | c | c   }
    \toprule
    Library & abort signals & segfaults & runtime error & inconsistent \\
    \midrule
    PyTorch & 0 & 2 & 5 & 3 \\
    TensorFlow & 24 & 2 & 0 & 1 \\
    \midrule
    Total  & 24 & 4 & 5 & 4 \\
    \bottomrule
\end{tabular}
}
    \vspace{-10pt}
\end{table}


\subsection{RQ2: Bug Discovery}
\label{subsec:rq2}

We conducted bug discovery~\cite{titanfuzz, pham2019cradle, fuzzgpt, wang2022eagle, wang2022eagle} on the latest versions of PyTorch (v2.2.1) and
TensorFlow (v2.15) by the time of writing. Note that due to the lack of
automated tools for bug analysis, confirming each bug requires significant human
effort. Therefore, we only analyze the fuzzing results of \tool. However, to
compare with other fuzzers, we examine if the bugs detected by \tool exist
in the older PyTorch and TensorFlow versions tested by other fuzzers.
If so, it implies that those fuzzers were unable to identify these bugs
(otherwise they would have been fixed).\footnote{At this step, we also search
for bug information in the community (e.g., PyTorch Forums) to make sure they
have not been reported by others before.}

We report the bug discovery findings in \T~\ref{tab:bugs}. On TensorFlow, \tool
identified 27 bugs, with 4 already fixed and 15 replicated by the developer 
but still under investigation.
Importantly, 16 bugs exist in TensorFlow (v2.10) which has been tested by
TitanFuzz and FuzzGPT. Nevertheless, neither TitanFuzz nor FuzzGPT were able to
detect them. Meanwhile, 16 bugs exist in TensorFlow (v2.6), which has been
tested by IvySyn. On PyTorch, \tool discovered ten bugs, with 4 already fixed 
and 4 replicated by the developer but still under investigation. 
Among these, eight bugs exist in
PyTorch (v1.12) which was tested by TitanFuzz and FuzzGPT, while six bugs
already exist in PyTorch (v1.11) tested by IvySyn. The above results demonstrate
that \tool not only efficiently discovers bugs but also identifies long-standing
bugs.

\parh{Bug Characteristics.}~\tool\ uncovers 37 bugs in total (the last row in
\T~\ref{tab:bugs}), which can be categorized into four types: abort signals,
segfaults, runtime errors, and inconsistent output. Runtime errors include
\texttt{INTERNAL ASSERT FAILED}, \texttt{MKL FFT error}, and \texttt{cuFFT
error}. We calculate the number of bugs for each category, whose results are in
\T~\ref{tab:bugTypes}. Most of the bugs discovered by \tool are due to the lack
of relevant validation checks in APIs. Since we test various possible edge cases
over input variables, most APIs tend to trigger abort signals or segfaults
directly when encountering unexpected inputs, resulting in fewer cases of
inconsistent outputs. Among all found bugs, 28 cause crashes. As noted in DocTer~\cite{xie2022docter}, “Despite receiving invalid inputs, DL API functions should not crash. Instead, they are expected to handle such inputs gracefully (e.g., through throwing an exception).” Thus, aligned with related works’ focus (e.g., DocTer and IvySyn), we deem these crash bugs as critical.

\parh{Transferability.}~While our edge cases were extracted from PyTorch, we
successfully discover 27 bugs in the latest version of TensorFlow. This
indicates that the edge cases we extracted have good transferability and can be
used across platforms. In fact, one could even interpret that the edge cases
extracted from PyTorch are ``more effective'' in uncovering bugs in TensorFlow
than that of PyTorch itself. We believe this is due to the fact that PyTorch has
been tested by many fuzzing tools and the community is actively fixing bugs
(still, we find ten bugs). Moreover, by transferring edge cases across
platforms, \tool\ for the first time enables a highly comprehensive fuzzing of
TensorFlow, which is hardly achieved by existing tools (most of those 27 bugs
are \textit{not} found by previous works).

\parh{Extracted Edge Case Types.}~The \tool\ is capable of extracting a wide variety of edge cases. Except the three types in Table~\ref{tab:edgeCase}, there are some other types: (1) edge cases related to multiple parameters, such as “\texttt{tensor_1} has a larger last dimension than \texttt{tensor_2}”; (2) some special parameter attributes related to program logic, such as “sparse”, “dense”, “conjugate”, and “contiguous” for tensor. (3) restriction between the attributes of a parameter, such as “\texttt{tensor_1} with input.shape[-2] $\textless$ input.shape[-1]”. LLMs employed by \tool\ can handle these cases properly.

\subsection{RQ3: Alternative LLMs}
\label{subsec:rq3}

In this RQ, we evaluate if \tool can effectively employ smaller LLMs for
edge-case based mutation. 
In our experiments, the general LLMs have demonstrated greater proficiency in understanding our tasks, particularly in creating edge cases for a target API.
Thus, we conduct experiments using two widely-used general
open-source LLMs, llama2-70b-chat and llama2-7b-chat~\cite{llama}. Recall we have
obtained a set of bug-triggering programs in \textbf{RQ2}; we collect the
prompts which are used to generate these programs, and then feed those prompts
into llama2-70b-chat and llama2-7b-chat. With the same prompts, we check if they
can also generate bug-triggering programs. We do not have powerful servers to
run llama2-70b-chat, and therefore, we use the APIs provided by
Replicate~\cite{replicate}. Replicate provides APIs capable of running
open-source models. We set the temperature to 0.5 for each prompt and repeat the
generation process three times for each prompt.

\begin{table}[!htbp]
    \centering
    \caption{Evaluating other LLMs.}
    \label{tab:diffLLM}
    \vspace{-5pt}
    \resizebox{0.8\linewidth}{!}{
    \begin{tabular}{c | c | c | c   }
    \toprule
    Target Library & ChatGPT-3.5 & Llama-2-70b-chat & Llama-2-7b-chat \\
    \midrule
    PyTorch & 10 & 3 & 2 \\
    TensorFlow & 27 & 10 & 4 \\
    \midrule
    Total & 37 & 13 & 6\\

    \bottomrule
\end{tabular}
}
\end{table} 

The results are in \T~\ref{tab:diffLLM}. Using llama2-70b-chat, we can still
discover three bugs on PyTorch and ten bugs on TensorFlow, respectively.
llama2-7b-chat results in discovering two bugs on PyTorch and four bugs on
TensorFlow. Interestingly, llama2-7b-chat discovers two bugs on TensorFlow that
are not found by llama2-70b-chat. Upon further analysis, we find that
llama2-7b-chat exhibits more uncertainty in generating complex tensors. This
increased uncertainty occasionally leads to correctly generating bug-triggering
programs.

\subsection{RQ4: Ablation Study}
\label{subsec:rq4}

%

\begin{figure*}[!htbp]
    \centering
    \includegraphics[width=0.9\linewidth]{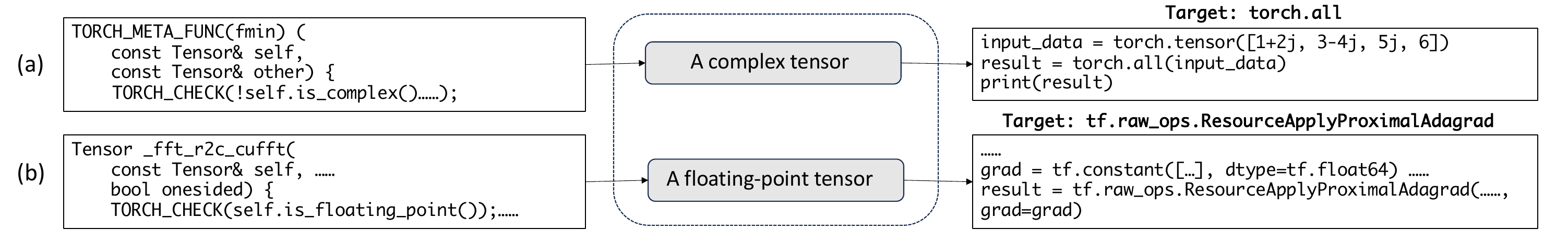}
    \vspace{-10pt}
    \caption{The case study.}
    \label{fig:caseStudy}
\end{figure*}

In \tool, each test case is formed by conducting two tasks: initial program
generation and edge case-based mutation. To analyze the contribution of each
component to bug discovery, we analyze the bugs discovered by each component.
We find that \textit{all} bugs are discovered through the edge case-based
mutations. This is unsurprising, because the API list we select is aligned with
both TitanFuzz and FuzzGPT. TitanFuzz and FuzzGPT have already tested these APIs
and discovered some rather easy-to-trigger bugs. Given \tool's initial program
generation is mainly to offer a program invoking the target APIs and thus
expanding API coverage, it is expected to not discover any new bugs. 

On the other hand, the results show the effectiveness of our edge case-based
mutation. Although these APIs have already been tested by other tools, our
edge case-based mutation approach still detected ten and 27 bugs respectively in
the latest versions of PyTorch and TensorFlow (where most of them also exist in
older versions tested by previous fuzzing tools). This illustrates that our tool
effectively benefits from diverse edge cases, consequently leading to the
discovery of more bugs.

\subsection{Case Study}

We present case studies to better illustrate the insights of \tool.
\F~\ref{fig:caseStudy} presents two examples to show how \tool\ extracts
context-free edge cases from the source code and successfully triggers bugs. In
\F~\ref{fig:caseStudy}(a), within \texttt{TORCH_META_FUNC}, \texttt{TORCH_CHECK}
checks if the variable with the tensor type deems a complex tensor. Accordingly,
we extract an edge case of tensor type: \texttt{complex tensor}. Since the input
parameter of API \texttt{torch.all} is also of tensor type, we use the extracted
edge case to fuzz it, by setting its input parameter as a complex tensor. This
way, we discover an inconsistent output bug in PyTorch (v1.12) running on GPU
and CPU, due to that \texttt{torch.all} does not consider the case of complex
tensor. Given that \texttt{torch.all} is widely used, our finding received
immediate attention from developers and was set as \textit{high priority} for
fixing.

In \F~\ref{fig:caseStudy}(b), we extract edge cases from PyTorch and apply them
in fuzzing TensorFlow; this shows the transferability. Within
\texttt{_fft_r2c_cufft}, \texttt{TORCH_CHECK} checks if the variable with tensor
type deems a floating-point tensor. Similarly, we extract an edge case of tensor
type: \texttt{floating-point tensor}, and use the extracted edge case to fuzz
\texttt{tf.raw_ops.ResourceApplyProximalAdagrad}. We find that this API has a
severe core dumped issue in handling floating tensors. 

It is difficult for previous fuzzing tools to detect these two bugs. TitanFuzz
applies basic mutation operators to mutate programs based on a scheduling
algorithm. However, it lacks guidance for triggering edge cases, making it
infeasible to trigger those two bugs. 
Similarly, for FuzzGPT to discover these bugs, three prerequisites must be met:
first, there must be bug codes with similar root causes on the Internet. Second,
FuzzGPT needs to properly crawl and collect these related bug codes. Third,
FuzzGPT must correctly extract the root cause from these bug codes and apply it
to test \texttt{tf.raw_ops.ResourceApplyProximalAdagrad}. However, these two
bugs already existed in PyTorch (v1.12) and TensorFlow (v2.10), which have been
tested by FuzzGPT; this indicates that FuzzGPT overlooked these bugs. Overall,
the aforementioned three conditions are uneasy to meet in practice, making it
difficult for FuzzGPT to discover these bugs. In contrast, \tool\ manifests two
advantages over FuzzGPT: (1) \tool does not depend on information (bug codes)
from the Internet; instead, it \textit{only} needs the source code of DL
libraries. (2) It can comprehensively test APIs, uncovering more edge cases that
developers may not have considered nor reported online.

IvySyn manually designs mutators through the analysis of known 240 CVEs. This is
a substantial effort, yet hard to be comprehensive. Particularly, for tensor
types, IvySyn features a pool of tensor mutations containing tensors with large
positive and negative values, tensors with empty shapes, and tensors containing
random dimensions. Nevertheless, it does not include cases of complex tensors
and floating-point tensors. As a result, IvySyn fails to find these two bugs,
although these bugs already exist in PyTorch (v1.11) and TensorFlow (v2.6) that
have been tested by IvySyn. This illustrates the inherent challenge faced by
manual efforts: it is difficult to be comprehensive and also keep up with the
rapidly evolving DL libraries. In contrast, \tool\ automatically extracts edge
cases from the source code, not only eliminating the need for expert experience
but also capturing a more comprehensive set of edge cases.





%% file: discussion.tex
\section{Discussion}
\label{sec:discussion}

\parh{Extension.}~Holistically, \tool\ leverages the reasoning and generation
abilities of LLMs to achieve comprehensive fuzzing of DL libraries. In this
regard, we envision several promising directions to extend \tool. First, we can
further improve the reasoning ability of LLMs by incorporating more
domain-specific knowledge. For example, we can leverage the existing knowledge
in the form of API documentation, code comments, and bug reports to guide the
reasoning process. As a common practice, this rich information can be dumped into
local vector databases, and then be used to guide the reasoning process~\cite{xie2022docter, li2020documentation}.
Second, we can enhance the generation ability of LLMs by incorporating more
advanced program synthesis techniques. Recent advances in LLM-based program
synthesis have shown promising results in generating programs under various
scenarios~\cite{liu2024your, wei2023copiloting, austin2021program,
zhang2023planning, sobania2022choose, li2024guiding, jain2022jigsaw}.
Nevertheless, we clarify that the benefit of enhancing generation ability may be
marginal, given that when preparing the initial test programs, we often do not
need to synthesize complex programs. 

\parh{Threat to Validity.}~One potential threat to the validity of our study is
the generalization of our findings. While we have demonstrated the effectiveness
of \tool\ on two popular DL libraries, TensorFlow and PyTorch, it is unclear
whether \tool\ can be generalized to other DL libraries. To mitigate this
threat, we illustrate the high transferability of \tool\ by showing that its
extracted edge cases from PyTorch can be effectively used to test TensorFlow.
This suggests that \tool\ can be generalized to other DL libraries. 

Another potential threat is the reproducibility of our findings. LLMs can be
sensitive to the training data, the model architecture, and the hyperparameters.
Particularly, commercial LLMs like GPT-3.5 only offer remote APIs which might
undermine its reproducibility. To mitigate this threat, we have made our
artifact available~\cite{dfuzz}, and we have assessed the performance
using alternative open-source LLMs. From another perspective, we observe that a
certain level of randomness is helpful to fuzz (e.g., our findings in
\S~\ref{subsec:rq3}).